# A Surgery-Detection Two-Dimensional Panorama of Signal Acquisition Technologies in Brain-Computer Interface

Yike Sun[+,1], Xiaogang Chen[+*,2], Bingchuan Liu[1], Liyan Liang[3], Yijun Wang[4], Shangkai Gao[1] and Xiaorong Gao[*,1]

[+]Y.S. and X.C. contributed equally to this work.
[*]To whom correspondence may be addressed.
(Email: chenxg@bme.cams.cn or gxr-dea@mail.tsinghua.edu.cn)

[1]Department of Biomedical Engineering, Tsinghua University, Beijing, 100084, China.
[2]Institute of Biomedical Engineering, Chinese Academy of Medical Sciences and Peking Union Medical College, Tianjin, 300192, China.
[3]Center for Intellectual Property and Innovation Development, China Academy of Information and Communications Technology, Beijing, 100161, China.
[4]Institute of Semiconductors, Chinese Academy of Sciences, Beijing, 100083, China.

## Abstract

Brain-computer interface (BCI) technology is an interdisciplinary field that allows individuals to connect with the external world. The performance of BCI systems relies predominantly on the advancements of signal acquisition technology. This paper aims to present a comprehensive overview of signal acquisition technologies in BCI by examining research articles published in the past decade. Our review incorporates both clinician and engineer perspectives and presents a surgery-detection two-dimensional panorama of signal acquisition technologies in BCI. We classify the technologies into nine distinct categories, providing representative examples and emphasizing the significant challenges associated with each modality. Our review provides researchers and practitioners with a macroscopic understanding of BCI signal acquisition technologies and discuss the field's major issues today. Future development in BCI signal acquisition technology should prioritize the integration of diverse disciplines and perspectives. Striking a balance among signal quality, trauma, biocompatibility, and other relevant factors is crucial. This will promote the advancement of BCI technology, enhancing its efficiency, safety, and reliability, and ultimately contributing to a promising future for humanity.

## Brain-computer interface

Hans Berger recorded the first electroencephalogram (EEG) signals in 1924 using clay electrodes on a 17-year-old college student with cranial defects, thereby pioneering a scientific technique for observing human brain activity [1,2]. Over the years, researchers have strived to integrate human brain signals with computer systems, leveraging the advancements in computer technology that employ electrical signals for communication. Jacques Vidal first proposed the concept of the brain-computer interface (BCI) in 1973 [3]. Since then, the technology has undergone remarkable progress [4-7], leading to the development of various BCI systems and expanding the definition of BCI technology. The first international conference in 1999 defined BCI as "a communication system that does not rely on the brain's normal output pathways of peripheral nerves and muscles" [8]. In 2012, researchers defined BCI technology as "a new non-muscular channel" for interaction [9]. Furthermore, in 2021, researchers introduced the concept of generalized BCI, which is defined as "any system with direct interaction between a brain and an external device" [10].

Figure 1 illustrates a schematic representation of a typical Brain-Computer Interface (BCI) system is illustrated. The components of BCI systems can be categorized into four main parts: signal acquisition, processing, output, and feedback. The signal acquisition component of a BCI system detects and records brain signals, which play a crucial role in determining the overall performance of the system. The processing component analyzes the recorded brain activity by utilizing specialized methods and algorithms to interpret the participant's intended action. The output component aims to execute the participant's intended action, typically achieved through the use of a robotic arm or speller using the processed information from the previous component. The feedback component informs the participant about the computer's interpretation of their intended action and conveys the final execution results through various sensory forms, including visual and auditory feedback. This allows for adjustments and supports closed-loop design.

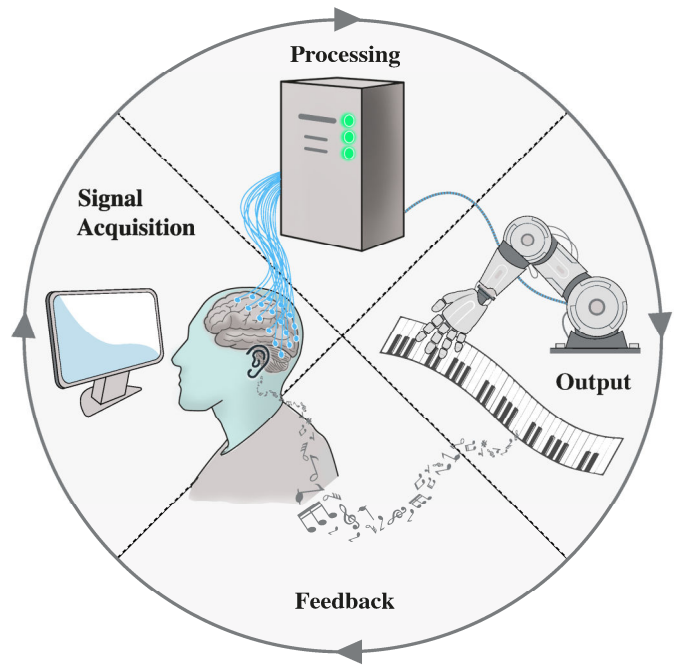

**Figure 1.** System structure of a typical BCI. It includes four parts: signal acquisition, processing, output, and feedback.

## Classification of signal acquisition technology for BCI

According to Jacques Vidal, a BCI is a device that utilizes EEG signals [3]. However, the diversification of signal acquisition methods due to technological advancements has made it challenging to precisely categorize BCI signal acquisition technologies. While the majority of researchers classify BCIs as non-invasive or invasive depending on the requirement of surgery, some have attempted to refine this categorization. In 2020, He et al. proposed a classification of flexible electrodes, categorizing them as non-intrusive, intrusive, and semi-intrusive [11]. This classification takes into account the degree of invasiveness and intrusion of the electrodes into the user's body. In 2021, Eric et al. classified the signal acquisition technology of BCIs into three categories: non-invasive, embedded, and intracranial [12]. This method considers the sensor's location relative to the brain and the degree of invasiveness.

The rapid development of science and technology has increased interest in BCIs, establishing it a popular and rapidly expanding field in recent years. This trend has attracted a significant number of scientists, engineers, and entrepreneurs, which has fostered innovation and advancements in BCI technology. The emergence of new technologies has brought new ideas and challenges to the traditional categorization of BCIs. Furthermore, progress in materials science and clinical medicine have also presented new opportunities in BCIs. To facilitate interdisciplinary communication and collaboration among researchers in various fields, we have meticulously organized existing works and proposed a novel and diverse perspective: the surgery-detection two-dimensional panorama of signal acquisition technologies in BCI. This panorama provides a comprehensive overview of current and potential BCI technologies, highlighting their respective advantages, disadvantages, challenges, and opportunities. It will serve as an invaluable reference and tool for future research and development endeavors in the field of BCIs.

## A two-dimensional view of BCI signal acquisition technologies

The existing classification methods predominantly adopt a single perspective when addressing the problem, such as focusing on the degree of invasiveness or the sensor's location. However, the prevailing view that trauma is synonymous with the degree of implantation is being challenged by emerging technologies. For instance, surgical enhancements to enhance propagation or the implantation of electrodes via vascular interventions offer alternative perspectives. Given the ongoing evolution of BCI technology and the increasing need for interdisciplinary collaboration, it is essential to embrace a multi-perspective classification approach. This approach enables us to gain a comprehensive understanding of the current state of BCI technology and facilitates the formulation of a potential roadmap for future advancements.

The design of BCI systems predominantly involves two key holders: clinicians and engineers. Clinicians are primarily involved in the surgical design aspect, focusing on addressing surgical trauma. On the other hand, engineers play a crucial role in signal acquisition and prioritize the optimal functioning of the sensor. This study presents a rigorous and comprehensive approach to evaluate the characteristics of various BCI signal acquisition techniques. In contrast to prior studies that focused on a singular aspect or dimension of BCI signal acquisition, we analyze the issue from two distinct perspectives: the surgical and sensor dimensions. Together, these two dimensions constitute a classification model that thoroughly encompasses the surgical aspect and the variances arising from diverse sensor operation modes.

Through the utilization of this model, we can simultaneously enhance guidance for surgery and sensor design, while comprehending the strengths, weaknesses, and potential risks associated with different BCI technologies. Moreover, we can ascertain the trade-offs and challenges involved in selecting and designing an optimal BCI signal acquisition technique for a specific application or user group. In the following sections, we will expound upon the characteristics of these two dimensions and provide a detailed comparison and evaluation of different BCI signal acquisition techniques based on this model. It is important to note that our objective is not to utilize two totally orthogonal vectors as a means to encompass all technologies. Rather, we aspire that this multi-perspective approach will enhance readers' comprehension of brain-computer interface technology and provide guidance for future advancements.

## Surgery dimension: invasiveness of procedures

The dimension of surgery significantly influences the feasibility of employing these techniques. This dimension is primarily classified from the clinician's perspective and refers to the invasiveness of the surgical procedure involved in signal acquisition techniques. It encompasses three levels: non-invasive, minimal-invasive, and invasive, as depicted in Figure 2a.

The classification of the three dimensions is contingent upon the level of invasiveness exhibited by the procedure. A technique is categorized as 'non-invasive' when the surgical operations involved in the signal acquisition are performed without causing anatomically visible trauma to the subject, typically limited to wounds at the micron level and above. On the other hand, a technique is labeled as 'minimal-invasive' if it results in anatomically visible trauma at the micrometer level and above during the surgical procedure, but without affecting the brain tissue. Finally, a technique is deemed 'invasive' if it leads to anatomically visible trauma at the micron level and above, specifically within the brain tissue, during the necessary surgical procedure for signal acquisition.

Across the three levels of the surgery dimension, ranging from non-invasive to invasive procedures, the extent of surgical trauma progressively escalates. Similarly, the ethical implications associated with the respective signal acquisition technologies also increase gradually. Concurrently, as surgical risks amplify, there is a heightened demand for optimal medical conditions, consequently leading to a gradual escalation in the challenges associated with its implementation. The majority of non-invasive techniques do not necessitate constant clinical supervision, whereas most minimally invasive procedures require the involvement of neurology or neurosurgery specialists, and nearly all invasive techniques mandate direct intervention from seasoned neurosurgeons. These factors must be considered when evaluating the feasibility and clinical applicability of BCI signal acquisition techniques.

## Detection dimension: operating location of sensors

Detection dimension is a crucial aspect of BCI technology, approached from an engineering perspective, which is directly related to operating location of sensors. Additionally, this dimension is directly linked to the theoretical upper limit of signal quality achievable with this technology, as well as to biocompatibility risk and other indicators. As illustrated in Figure 2b, the dimensions are divided into three levels: non-implantation, intervention, and implantation.

The classification in detection dimension is primarily determined by the sensor's location during operation. A BCI technology is categorized as 'non-implantation' if the signal is acquired through a sensor in vitro. On the other hand, 'intervention' is adopted from clinical medicine procedures in the field of interventional cardiac catheterization [13,14]. Sensors used in interventions leverage naturally existing cavities within the human body, such as blood vessels, to function without causing harm to the integrity of the original human tissue. if the signal source relies on a sensor located within the body's natural cavity, it is classified as an 'intervention' technology. In cases where the signal is collected from an implanted sensor within human tissue, the technology is labeled as an 'implantation' technology. We propose this dimension as a helpful guideline for engineers in the design of sensors.

The theoretical upper limit of signal quality depends on the distance from the signal source and the type of interlayer. For example, detecting brain activity is analogous to listening to a chorus of students in a classroom. Non-implantation methods are like listening from outside the building, where only a large-scale sum of neuronal activity can be heard amid many sources of noise. Intervention methods are like listening in the corridor, where more information can be collected and less interference can be encountered than non-implantation methods, but the noise from the internal environment (tissue fluid, blood, etc.) is still significant. Implantation methods are like listening inside the classroom, where the signal is clearer and less interfered. However, signal quality also depends on other factors, such as spatial resolution and sensor material properties. The above discussion only considers the maximum signal quality that can be achieved at a single point.

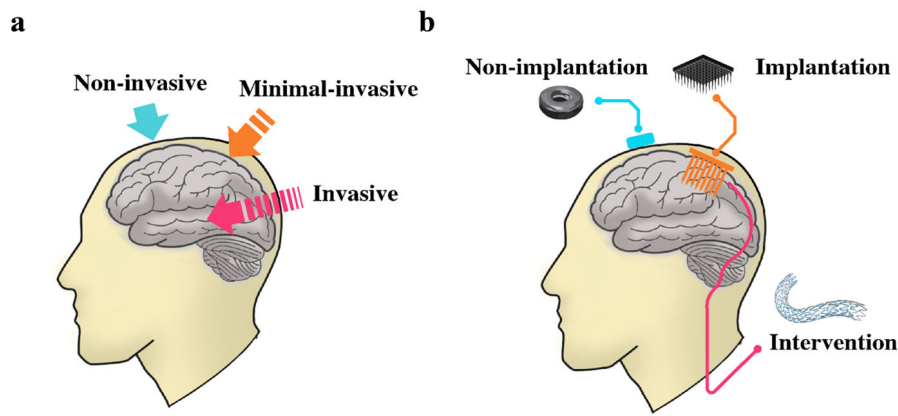

**Figure 2.** Classification of BCI signal acquisition technologies. (a) is the classification diagram of the surgery dimension, which includes three levels: non-invasive, minimal-invasive, and invasive. (b) shows the classification diagram of the detection dimension, which includes three levels: non-implantation, intervention, and implantation.

## Relationship of Detection dimension with signal

Current research in BCI signal acquisition has primarily centered on the detection aspect. Various methods, including implantation, intervention, and non-implantation techniques, are often considered in competition. Nonetheless, these three approaches operate within distinct anatomical locations, leading to limitations in their convergence. It is plausible that they will continue to develop independently.

Implantable sensors, owing to their close proximity to nerve cells, excel in capturing precise, high-frequency signals such as Local Field Potentials (LFP) and SPIKE signals. The latter denotes the firing of an individual neuron, manifesting a frequency exceeding 300 Hz. Conversely, LFP signals typify synchronized oscillations from small clusters of neurons within the 300 Hz frequency spectrum. Confrontationally, intervention techniques lack direct contact with neurons due to their isolation from soft tissues, including meninges and blood vessels. This isolation complicates the retrieval of SPIKE signals, which are dominated by LFP signals. Non-implantation techniques are confined to acquiring large-scale synchronized discharges with frequencies of 100 Hz or less, such as EEG, particularly when investigated in vitro.

Recent advancements in neuroscience underscore that electrophysiological signals like SPIKE, LFP, and EEG are not mere superimpositions; rather, they exhibit noteworthy scaling effects [15]. For instance, implantable sensors procure more precise signals; however, relying on a limited neuronal population might not faithfully represent complex cognitive processes within the human brain. Thus, the implantation of diminutive electrodes for signal acquisition in domains like affective computing could compromise signal fidelity and the overall efficacy of biometric applications. On the contrary, extensive signals (e.g., EEG) struggle to pinpoint specific regions of motor control due to the potential masking of subtle signal nuances resulting from broad neuronal firing. Consequently, systems employing finer-scale signals might demonstrate superior performance in tasks of this nature.

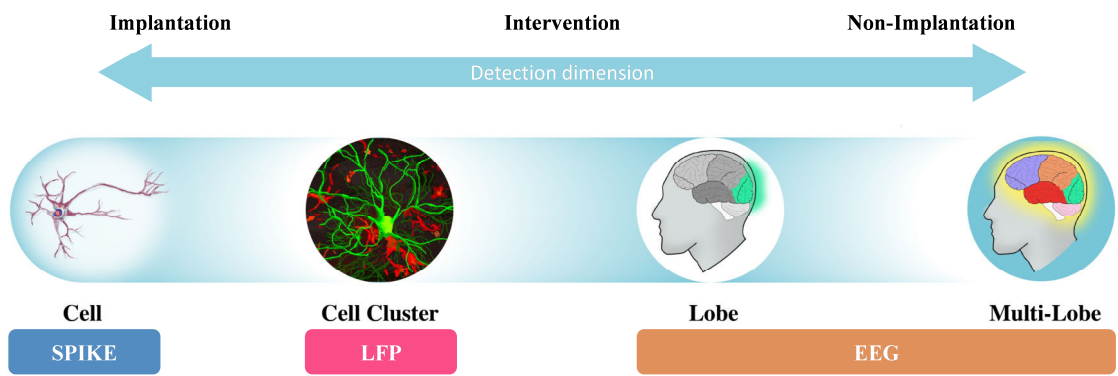

**Figure 3.** Schematic depiction of the Detection dimension concerning the acquired signal. Implantation technology primarily records signals at both the cellular and cell cluster levels. Intervention techniques gather signals from cell cluster clusters, alongside background signals originating from various brain regions. Conversely, non-implantation techniques yield signals either from the entire brain or specific brain regions. Specifically, the SPIKE signal corresponds to the electrical activity on a cellular scale, while LFP pertains to the electrical signal originating at the level of cell clusters. In contrast, EEG corresponds to the electrical signal originating from a singular brain lobe or multiple brain lobes simultaneously.

## The surgery-detection two-dimensional panorama of signal acquisition technologies in BCI

The surgery-detection two-dimensional panorama categorizes all BCI signal acquisition technologies into nine distinct categories (3x3), based on the aforementioned dimensions. This method furnishes a comprehensive and enlightening overview of the diverse signal acquisition techniques employed in BCI. Such an overview proves valuable in the selection and development of appropriate techniques for varying applications.

To facilitate a comprehensive and meticulous classification analysis, an exhaustive search was conducted in the Web of Science database for articles pertaining to BCI or brain-machine interfaces (BMI). Articles indexed in the Science Citation Index (SCI), Emerging Sources Citations Index (ESCI), and Social Sciences Citation Index (SSCI) from 2012 to August 2022 were included. A thorough screening of the articles was conducted to exclude papers that focused solely on algorithms or reviews. Our filtering criteria were centered on the presence of signal acquisition experiments. Articles containing such experiments were included in the total count, while those lacking them were excluded, resulting in a total of 6,679 articles obtained. These articles were classified utilizing our proposed classification model. Due to the space limitations, not all of these articles are shown in the reference section.

The outcomes of the classification, along with their corresponding proportions, are presented in Figure 4. Currently, the dominant approach in BCI/BMI research is non-invasive non-implantation technology, constituting approximately 85.87% of the studies. In contrast, the utilization of non-invasive intervention technology, minimal-invasive non-implantation technology, and minimal-invasive intervention technology is still in its nascent stages, collectively representing only 0.13%, 0.02%, and 0.06% of the studies, respectively. Minimal-invasive implantation technology, characterized by numerous clinical studies, primarily focuses on experiments involving epilepsy and paralysis patients, accounting for 4.84% of the studies. Lastly, invasive implantation technology, predominantly employed in animal and patients with paralysis studies, accounts for 9.08% of the studies.

This section, dedicated to the classification of technology in existing articles, serves the purpose of providing readers with a rapid comprehension of the present research landscape within the field. It is important to note that the prevalence of signal acquisition technologies over time does not inherently dictate the quality or effectiveness of their deployment. Looking ahead, significant shifts are anticipated in the distribution of each technology's prevalence. Non-invasive, non-implantable technologies are expected to witness a gradual decline in deployment complexity due to ongoing advancements in clinical and materials domains. In parallel, the proportion of other technologies is projected to experience a noteworthy upsurge.

Each BCI signal acquisition technology encompasses unique application scenarios, along with corresponding advantages and disadvantages. The subsequent sections offer a comprehensive survey of the customary techniques within each classification category.

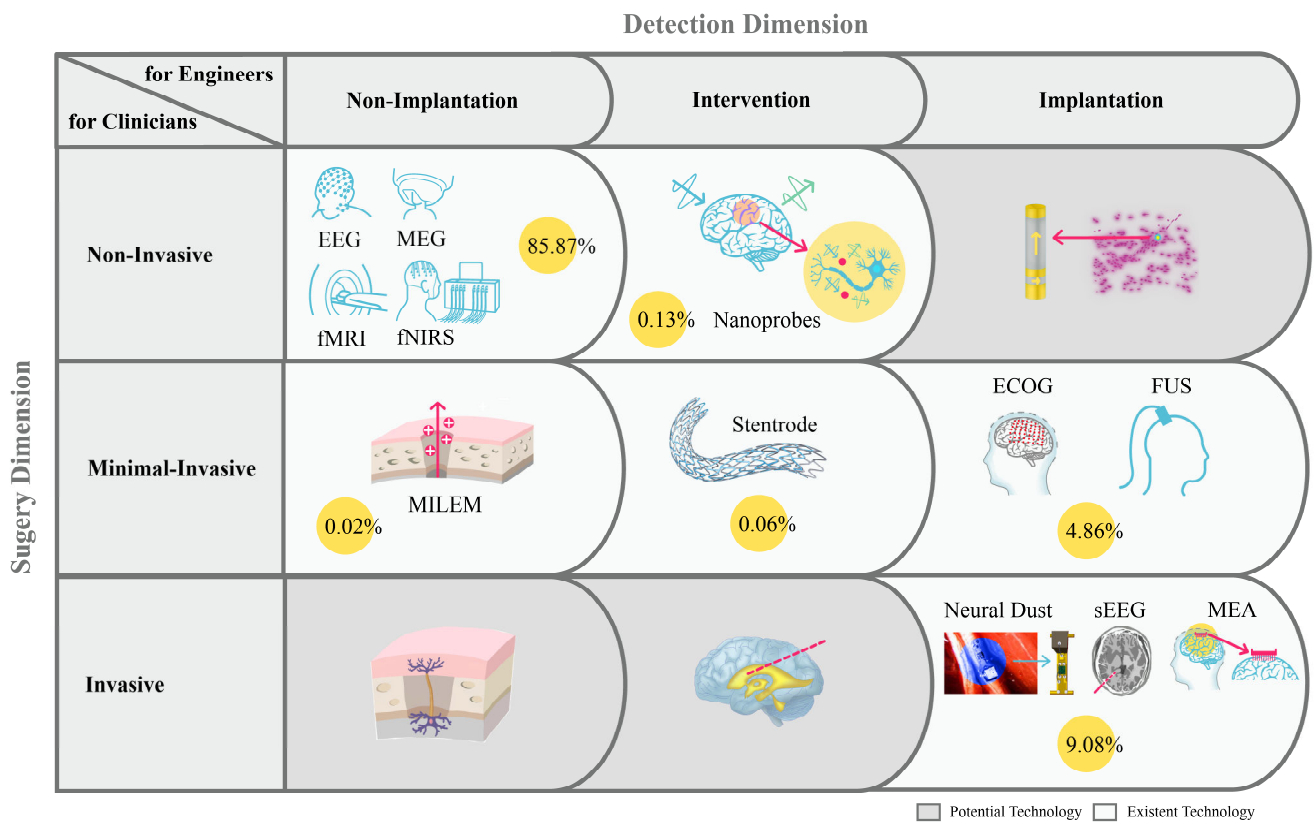

**Figure 4.** The surgery-detection two-dimensional panorama of signal acquisition technologies in BCI. It includes two dimensions, totaling 9 (3x3) technology types. We surveyed 6679 research articles in this decade and obtained the proportion of each technology. They are non-invasive non-implantation technology (85.87%), minimal-invasive non-implantation technology (0.02%), non-invasive intervention technology (0.13%), minimal-invasive intervention technology (0.06%), minimal-invasive implantation technology (4.84%) and invasive implantation technology (9.08%). Grids with white backgrounds in the figure represent existing technologies and grids with gray backgrounds represent potential ones, which will be detailed below.

## Non-Invasive Non-Implantation Technology

Non-invasive non-implantation technologies are for signal acquisition due to their portability and applicability and applicability without implants. Electromagnetic and blood flow signals are two main categories of non-invasive non-implantation methods. The electromagnetic signal category includes Electroencephalogram (EEG) and Magnetoencephalogram (MEG). EEG is widely used for its low cost and ease of use, which has an excellent time resolution [16,17]. However, its signal quality can be degraded by tissues [18,19] and other bioelectric interference [20-22]. Most EEG devices uses wet electrodes [23-25], but dry electrodes are being explored [26-28], such as microneedle [29-33] and direct-contact electrodes [34-37]. MEG, like EEG, has an excellent temporal resolution [38-40] but more channels [41]. MEG equipment usually rely on SQUID as the core [42], which requires exceptionally high magnetic shielding [17,43,44] and liquid nitrogen for cooling, which is costly [45]. Recently, optically pumped magnetometer (OPM) devices have been proposed to address these issues to to some extent [46,47]. Moreover, the signals obtained by EEG and MEG are somewhat complementary [17], which suggests the possibility of combining them [48]. It is noteworthy to mention that the electromagnetic signals obtained through non-invasive non-implantable techniques constitute the vague outcome of collaborative neuronal activity involving a substantial number of neurons. These signals predominantly manifest within the realm of low frequencies.

Blood flow signals are indirect measures of neuronal activity, unlike electromagnetic signals, so blood flow signals have a natural disadvantage in time resolution [49]. Three technologies that detect changes in blood flow signals are functional Near-infrared Spectroscopy (fNIRS), functional Transcranial Doppler (fTCD), and functional Magnetic Resonance Imaging (fMRI). fNIRS uses near-infrared light to detect changes in oxygen content in the blood [50,51], offering high spatial resolution [52,53]. However, the skull is also a poor conductor of near-infrared light [54], limiting the depth of information that can be received. At the same time, when the distance between the near-infrared emitter and the receiver is less than one centimeter, the collected signal is mostly derived from the skin [55]. fTCD uses ultrasound Doppler imaging, which has a high imaging rate [56-58]. Some researchers also suggest fTCD can be a high-resolution resolution replacement for fNIRS [59]. However, fTCD requires couplant, which reduces its convenience. fMRI has a very high spatial resolution, up to a fraction of a millimeter [60], but traditional imaging has a high delay [61,62]. Although some fast-imaging methods are available [63], the temporal resolution of fMRI still falls behind other BCI signal acquisition techniques [64]. Blood flow signal techniques are commonly used in multimodal BCI studies for good electromagnetic compatibility [65-75].

## Minimal-Invasive Non-implantation Technology

The minimal-invasive non-implantation technologies aim to solve the technical obstacles of non-implantation technology by using minimally invasive surgery. The main obstacle for recording EEGs, as Hans Berger struggled for 27 years is the interference from various tissues in the human body [76], especially the skull, which has low electrical conductivity [18]. To enhance signal transmission, researchers developed the Minimally Invasive Local-Skull Electrophysiological Modification (MILEM) technology [77], which uses ultrasonic vibration to create a small hole in the skull, improving the electric field distribution on the scalp. MILEM preserves high time resolution and significantly increases the signal-to-noise ratio compared to conventional EEG methods. Although this technology does not avoid surgery completely, it does avoid implantation. This tech can be a potential solution for future applications.

## Non-Invasive Intervention Technology

Intervention methods aim to place sensors in the natural cavity of the human body to avoid problems such as inflammation. Non-invasive intervention is a promising direction for BCI research for safety and universality. Research in this field can be divided into two categories based on the implanted cavity: blood vessel and ear canal. The blood vessel category uses nanoprobes to obtain the signal remotely, while the ear canal category records EEG. The nanoprobes method has been widely studied [78-80] as a medical imaging method, but only Neuro-SWARM3 used this method in the BCI signal acquisition [81], which utilizes nanoprobes functionalized with lipid coatings injected into the circulatory system [82,83]. However, there is no clear evidence for the signal quality and no report of in vivo experiments. In-ear EEG is a non-invasive method that uses an ear canal to record EEG. The device is similar to earplugs [84-86], and the signal characteristics are similar to the T7, and T8 leads in the 10-20 lead system [87]. However, there is a challenge in selecting ground and ref leads, with some studies placing them on the outer ear or scalp [86,88,89], compromising portability advantages. Alternatively, other studies have put them together in the ear canal [87], affecting signal quality.

## Minimal-Invasive Intervention Technology

Minimal-invasive intervention technology methods provide more accurate and deeper signals than non-invasive intervention methods. The stent-electrode recording array technique, or Stentrode, proposed by Oxley et al. in 2016, is a representative technology in this field, focusing on blood vessel category. It involves implanting a stent electrode array through minimally invasive surgery into the cerebral vein in the sulcus fold, using natural cavities to place sensors, and avoiding some immune responses [90,91]. Stentrode uses the venous sinus stenting surgical technique to navigate and insert the self-expanding scaffold electrode array into the target. The safety of the operation has been confirmed [92-94], and it has acceptable biocompatibility [95,96]. The signal bandwidth recorded by Stentrode can reach 226HZ [97], and most signals collected are derived from LFP near the implantation site [98]. However, the technology has some drawbacks, such as the complexity and risks of complications such as intracranial hemorrhage and thrombosis [99]. Furthermore, installing a signal transmitter under the clavicle [100] increases the cost of the operation. Stentrode's potential still requires

## Minimal-Invasive Implantation Technology

Minimal-invasive implantation technologies represent an emerging area of research in BCI development. This field encompasses two categories of technologies: acoustic and electrical. Acoustic signal technologies, such as focused ultrasound imaging (FUS), employ ultrasound as the signal transmission medium and implanted sensors to acquire and analyze brain signals. Unlike conventional ultrasound techniques such as fTCD, FUS is a minimally invasive neuroimaging technique that involves direct implantation of transmitter and transducer elements outside the dura to produce high-resolution signals with high sensitivity [101-104]. It mainly detects blood movement [105], which makes it useful for motion decoding studies [106,107]. However, FUS is still in the development, and its applications and limitations need further investigation.

In electrical signal technologies, subcutaneous EEG (sqEEG) is a long-term [108] wearable system that consists of a small subcutaneous implant and an outer part that provides power [109], which has similar signal characteristics to conventional EEG [110,111], and it offers advantages in motion artifact suppression. Currently, sqEEG is mainly used in epilepsy detection but holds significant potential in BCI research [109]. Electrocorticogram (ECoG), a method first proposed in the 1950s for epilepsy localization [112], has higher spatial resolution and bandwidth than EEG [113,114], and is less susceptible to EMG and EOG interference [115,116]. However, ECoG requires risky surgery for implantation [117], limiting its acceptance for non-therapeutic purposes. Nevertheless, ECoG-based BCI research is prominent, as most EEG-based BCI paradigms can be reproduced with ECoG, yielding better results [118-120]. ECoG also has considerable potential in speech and action decoding [121-123], shown by a 2019 study that directly converted neural activity into speech using ECoG signals [124]. As research progresses, these technologies may revolutionize the field of BCI and offer groundbreaking therapeutic applications.

## Invasive Implantation Technology

Invasive implantation technologies are a popular category in current research because of their spatial resolution and signal bandwidth advantages, which allow for excellent decoding operations [125]. They are utilized in BCI solutions such as BrainGate [126] and can be classified into cortical and depth signal classes. The cortical signal class mainly consists of Neuralink, Neural Dust, and intracortical microelectrode arrays (MEAs). Neuralink proposed a scalable high-bandwidth brain-computer interface platform in 2019 comprising high-density electrodes, an automated surgical robot, and a small implantable processing device [127]. Although Neuralink's solution has the potential to result in a high-quality stereo signal with less surgical trauma, the specific performance of the technology has not been thoroughly evaluated due to a lack of research reports. Neural Dust is a tiny sensor cluster that can monitor and stimulate neuronal activity in the brain or other body parts by ultrasound [128]. It is powered remotely by ultrasound without battery implantation [129] and can wirelessly transmit data for processing and analysis [130,131]. Although the Neural Dust has so far only been performed experimentally on peripheral nerves [132], it was originally designed to be able to performed as a central neural interface [130]. We still consider it a BCI technology.

The commonly used MEAs include the Utah Array [133] and Michigan probes [134]. MEAs are utilized in the decoding tasks of motion [135], vision [136], and speech [137]. In 2021, Willett et al. employed the BrainGate system to achieve a brain-to-text communication scheme by interpreting handwriting [138]. The primary signals consist of LFP (low-frequency) and SPIKE (high-frequency) signals. MEAs have been continuously used for several months to several years after implantation [139]. However, the SPIKE signal weakens shortly after implantation due to pin loss and frequency band reduction caused by the immune response. LFP signal components primarily serve as the basis for long-term MEAs [140,141]. Over the years, researchers have made significant advancements in the field of flexible MEA. For example, in 2019, Guan et al. proposed the Neural Matrix [142], which is an electrode developed using flexible silicon film transistors known for their excellent scalability. In 2020, Zhao et al. introduced ultra-flexible neural electrodes that utilize bio-dissolvable adhesive, facilitating long-term stable intracortical recording [143]. Some researchers have utilized carbon nanotubes in the development of flexible electrodes, which possess characteristics such as low toxicity and excellent electromagnetic compatibility [144]. In 2022, Zhou et al. presented an implantable electrode created from silk protein, which can avoid contact with tissues like blood vessels and effectively minimize the invasiveness of implantation [145]. The longest known MEA implanted is a neurotrophic electrode, a method of growing neurites into the electrode tip [146], which an implantation time is 13 years [147].

Depth signal class technologies, such as stereotactic electroencephalography (sEEG), Neuropixels, and fully implantable BCI, offer enhanced insights into the depth dimension of the signal. sEEG is a deep implantable signal acquisition technology with higher bandwidth, signal amplitude, and spatial resolution compared to standard EEG [148-150]. This makes it valuable for localizing epileptic lesions [151,152] and decoding speech [153] and motion [154]. Neuropixels is a type of fully-integrated silicon complementary metal-oxide semiconductor (CMOS) digital neural probe [155] that falls under the category of MEAs. It possesses the capability to simultaneously record cell-level signals at a highly reduced electrode scale [156], as well as exhibiting excellent performance in deep signal recording [157]. Due to its distinctive attributes, Neuropixels is regarded as a separate entity. In vivo animal experiments have extensively utilized Neuropixels [158], and in recent years, successful human experiments have also been conducted [157,159]. Notably, in 2021, researchers introduced Neuropixels 2.0, which enables more stable and prolonged recording compared to its predecessor [160]. Fully Implantable BCI is a technique that utilizes completely implanted electrodes and signal transmission units, effectively mitigating the risks associated with exposed wires and devices [161-163]. The most prevalent application of this technology is closed-loop BCI systems [164,165].

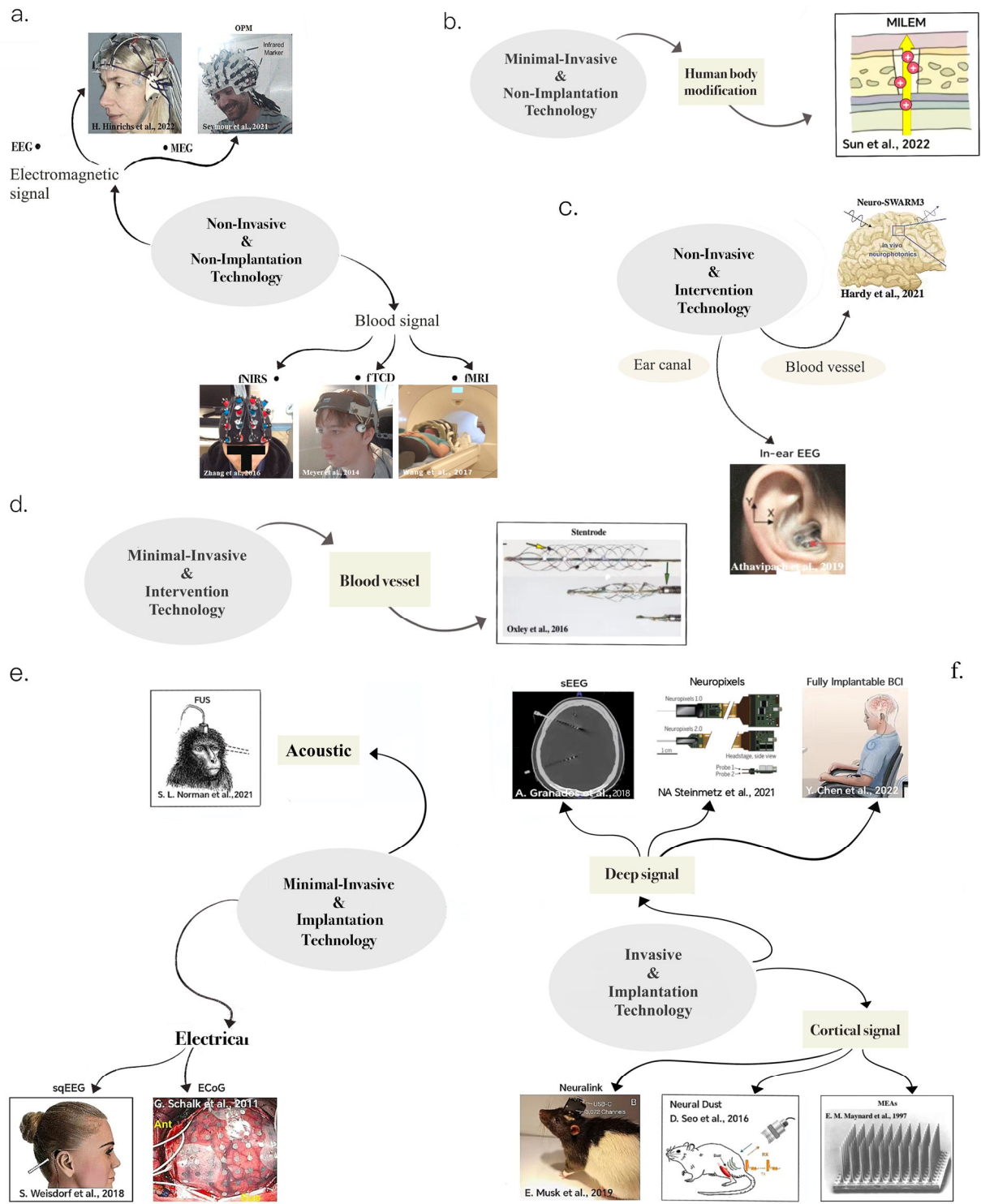

**Figure 5.** Representative technologies in each category. (a) Non-invasive non-implantation technology includes the electromagnetic signal category and blood signal category. The representative technologies of the electromagnetic signal are EEG [23] and MEG [44]. The representative blood signal technologies include fNIRS [51], fTCD [58], and fMRI [75]. (b) The main idea of minimal-invasive non-implantation technology is human body transformation, and MILEM [77] is the representative technology in this field. (c) Non-invasive intervention technology can be divided into the blood vessel category and ear canal category according to the type of implanted cavity. The representative technique in the blood vessel is Neuro-SWARM3 [81], and the representative technique of the ear canal is in-ear EEG [87]. (d) The representative minimal-invasive intervention technology is Stentrode [90], which places the sensor in the cerebral venous sinus through vascular navigation and belongs to the blood vessel category. (e) Minimal-invasive implantation technology includes acoustic and electrical categories. The representative technology of the acoustic category is FUS

[106]. The representative technologies of the electrical category are sqEEG [110] and ECoG [117]. (f) Invasive implantation technology is mainly concentrated in cortical and depth signal categories. The cortical signal category's representative technologies are Neuralink [127], Neural Dust [132], and MEAs [133]. sEEG [152], Neuropixels [160], and fully implantable BCI [161] belong to the depth signal category..

# Potentially feasible technology

### Non-Invasive Implantation technology: Tissue Penetration Nanorobot

The investigation of nanorobots has been stimulated by naturally occurring nanoscale mechanical structures, such as bacterial flagella and rotors [166]. Within the medical domain, nanorobots represent nanostructures with the capacity to perform surgical procedures, facilitate drug delivery, enable imaging, and conduct analysis [167,168]. While the majority of nanorobots are transported through the bloodstream and function within blood vessels, nanorobots with tissue-penetrating capabilities can access tissues inaccessible to blood by employing magnetic drilling and acoustic micro-cannon techniques [169,170]. In 2019, Jafari et al. effectively showcased the feasibility of introducing nanorobots into the brains of rats using magnetic drilling under the guidance of an external magnetic field [171]. This technology holds promise for applications involving the non-invasive implantation of BCIs. The envisioned scenario entails the injection of nanorobots into bloodstream of the brain, with an external magnetic field orchestrating their penetration into the cerebral cortex. Subsequently, these nanorobots could convey neural activity within the brain through ultrasound or near-infrared light.

### Minimal-Invasive Non-Implantation technology: Bone Tissue Replacement

The electrical impedance of the skull plays a crucial role in influencing the transmission of EEG signal. Thus, reducing skull impedance through minimally invasive non-implantation technology can enhance signal quality [77]. Achieving a decrease in skull impedance comparable to that of tissue fluid, thereby rendering the skull "electrically transparent," could potentially enable unobstructed observation of the brain's electrical activity in vitro. Bone tissue replacement surgery stands as a potential approach to achieving this goal. Research into artificial bone tissue and bone graft surgery has gained substantial traction [172,173]. Metals, inorganic non-metallic materials, polymers, and composite materials comprise the spectrum of artificial bone materials [174]. Identification of a material exhibiting both low electrical impedance and satisfactory biocompatibility could potentially pave the way for viable skull replacement surgery, consequently achieving electrical transparency of the skull. This technological advancement holds considerable promise for the advancement of BCIs. In the event that a forthcoming material is discovered, enabling the validation of this technical solution and facilitating unhindered interaction of ionic currents both within and outside the cranial cavity, it is anticipated that the attainable signal quality threshold will surpass that of the MILEM technique. This threshold should approximate or potentially fall below the level achieved by the ECoG method.

### Invasive Non-Implantation technology: living autologous neural device

The utilization of autologous living cells as a foundational substrate for constructing implantable electronic devices has been regarded as a promising avenue for circumventing immune and inflammatory reactions [175]. Currently, the application of this technology within the realm of BCI remains limited. In 2017, Serruya et al. pioneered the development of an autologous living-cell neural interface, employing neuronal clusters to establish a micro columnar architecture; notably, this interface is encased within a biodegradable hydrogel that naturally degrades in vivo [176]. Likewise, in 2021, Prox et al. devised a DBS device, wherein autologous neuronal cells and cardiomyocytes are enshrined within an agar gel shell [177]. This innovation facilitates the utilization of autologous cells to transmit signals directly from a targeted brain region externally, obviating the need for sensor implantation. Nevertheless, this approach necessitates invasive surgical implantation within the brain tissue, thus falling under the purview of invasive methods. It is important to acknowledge that despite the use of autologous cells, the human body may still mount an immune response to cells originating from distinct sites, underscoring the critical significance of judicious cell type selection within this research trajectory.

Invasive Intervention technology: Cannula Implantation in Brain Ventricular System

The ventricular brain system, which comprises four interconnected brain ventricles [178], is an integral component the brain. Extensive research has been conducted on the implantation of cannulas into brain ventricles for therapeutic or surgical procedures [179-181]. Moreover, studies have demonstrated the utility of electrical signals obtained from the ventricular brain system for BCI research [182]. Electrodes can be introduced into the ventricular brain system through a cannula implantation procedure, facilitating direct transmission of brain activity signals from the ventricles to external recording devices. Nevertheless, this procedure carries the potential risk of damaging brain tissue owing to osmotic pressure variations within the cerebrospinal fluid. Hence, this methodology is categorized as an invasive intervention technology. Nonetheless, considering that the neural tissue associated with consciousness resides on the dorsal aspect of the brain [183], while the ventricles are positioned within the ventral region, this method holds the potential to yield distinctive signals in contrast to conventional methodologies. These distinctive signals entail diverse application scenarios and signal characteristics.

## Conclusion and outlook

The realm of BCI technology confronts a multitude of challenges and limitations, particularly in the sphere of signal acquisition technology, which stands as the foundational cornerstone of any proficient BCI system. Distinctive attributes encompass invasiveness, spatial and temporal resolution, signal fidelity, cost considerations, usability, and safety across various signal acquisition methodologies. Therefore, a comprehensive analysis and comparison of prevailing signal acquisition technologies are imperative to comprehend their respective merits, drawbacks, and appropriateness for varied BCI applications. Within this manuscript, we undertake an exhaustive scrutiny of scholarly articles disseminated over the past decade, approaching BCI from the vantage point of signal acquisition. Introducing a novel and systematic framework, termed the "surgery-detection two-dimensional panorama," we adeptly structure and classify extant endeavors within the realm of BCI signal acquisition technology. This endeavor serves to furnish a lucid and succinct overview of the present state-of-the-art in BCI signal acquisition technology, while also delineating future trajectories and affording an invaluable instrument for interdisciplinary discourse and collaboration amongst BCI researchers.

Subsequent to the thorough investigation of diverse signal acquisition technologies, this article culminates by delving into the prospective pathways of BCI technology. Hinging on high-performance acquisition technology, the crux of propelling brain-computer interfaces toward practical applications lies in system-integrated solutions. To illustrate, through the establishment of stable signal acquisition and the formulation of scientifically grounded interaction paradigms, tangible enhancements in patients' well-being, augmentation of individual capacities, provisioning of enhanced daily conveniences, and the assurance of user safety can be viably realized. It is pertinent to acknowledge that the evaluations proffered in this discourse do not delineate the confines of application for individual technologies. Instead, we espouse the belief that all the technologies scrutinized harbor latent potential for further advancement and refinement. Nonetheless, certain technologies may manifest more pronounced advantages or disadvantages within specific domains or scenarios. By dispensing invaluable references and guidance, we aspire to kindle novel concepts and innovations within this exhilarating and promising arena of research and development.

Non-Invasive Non-Implantation technology heads for consumer electronics

The realm of BCI technology offers a plethora of potential applications in the realm of consumer electronics. These applications span various domains, including education [184], gaming [185,186], and communication [187,188]. Particularly, non-invasive, non-implantation technology stands out due to its remarkable social acceptance, evident cost-efficiency, and well-established technical maturity. Although it is acknowledged that this technology may exhibit lower signal quality compared to implantation technologies, its aptness for the target demographic of consumer electronics—comprising generally healthy individuals—is undeniable. This is especially relevant given the unlikelihood of such individuals pursuing surgical interventions for consumptive purposes. Hence, the present context strongly supports the preference for non-invasive non-implantation technology within the realm of consumer electronics. However, it is pertinent to acknowledge that challenges persist within consumer electronics applications, notably in cases where EEG is employed. Issues stemming from hair interference and device dimensions continue to pose hurdles. This calls for a dedicated focus on industrial research aimed at enhancing both electrodes and algorithms pertinent to non-invasive non-implantation technology.

Recent times have witnessed substantial strides in the practical deployment of non-invasive non-implantation BCI technology. Within the healthcare sphere, this technology has found valuable application in aiding stroke patients' motor rehabilitation [189,190]. The clinical validation of this application has rendered it a staple in clinical practice. Furthermore, the technology's real-time monitoring capability has been harnessed to gauge patients' anesthetic statuses during surgical procedures, thereby enabling dynamic anesthesia through closed-loop mechanisms [191,192]. This, in turn, has proven instrumental in mitigating the risks linked to anesthetic drug overdose. The convergence of non-invasive non-implantation neuromodulation technology has yielded products capable of effecting closed-loop sleep regulation, thereby enhancing sleep quality among patients [193,194].

Beyond the medical realm, this technology has manifested its utility in safety-critical operations. It has been instrumental in monitoring the cognitive states of specialized operators, thereby preempting conditions like severe fatigue [195]—particularly relevant to miners or truck drivers. The education sector also benefits from non-invasive non-implantation technology, employing them to assess group learning dynamics and bolster user concentration by tracking attention levels [196]. In the realm of entertainment, these interfaces facilitate users' meditation training and attainment of mind-flow states [197]. A noteworthy innovation comes in the form of virtual reality all-in-one devices, seamlessly integrating non-invasive non-implantation BCI technology to monitor brain states during gaming and enable brain-computer interaction [198,199]. Notably, brain-controlled mice founded on non-implantable brain-computer interfaces expand interaction possibilities for special populations, albeit at a slight performance trade-off compared to traditional mice.

In summary, the application landscape for non-invasive non-implantation BCI technology is broad and promising, especially within contexts emphasizing lightweight design, user-friendliness, and cost-effectiveness. This potential is particularly pronounced within consumer electronics and light medical applications.

## Individual equipment control in complex scenarios requires a reliable and stable signal source

BCI technology extends its utility beyond the realm of disability, demonstrating equal effectiveness for individuals without impairments. Through direct interfacing with personal devices like mechanical exoskeletons, BCI technology holds the potential to significantly augment the capabilities and survivability of individuals facing hazardous scenarios, such as those encountered in network-deprived environments like mines. This technology unveils novel prospects for navigating intricate settings like industrial construction, disaster relief operations, and exploratory ventures. Ongoing research in the field of BCI concentrates on the control of exoskeletons [200,201], unmanned aerial vehicles (UAVs/UGVs) [202,203], and prosthetic devices [204,205].

However, the acquisition of reliable BCI signals within intricate real-world situations remains a formidable challenge. Prevailing non-invasive techniques, frequently employed, suffer from suboptimal signal fidelity. The employed sensors often struggle to discriminate between brain-derived signals and other extraneous physiological noises. Conversely, implantation techniques directly within the cerebral cortex offer superior signal quality but are susceptible to damage and infections. The brain's immune response tends to encapsulate implanted devices over time, consequently degrading signal quality. Thus, the pursuit of minimal-invasive non-implantation, as well as minimal-invasive intervention methodologies, emerges as a plausible resolution. The placement of temporary electrodes within the cranial cavity, albeit external to the brain, emerges as a promising compromise between signal fidelity and safety. Another viable avenue involves the development of diminutive, more pliable, and less intrusive implants specifically designed for minimally invasive implantation techniques. The employment of advanced neural materials and electrode configurations that elicit minimal immune response from the brain presents the potential to facilitate the creation of safer, enduring implantation solutions.

Nevertheless, the ethical quandaries surrounding privacy, security, and identity necessitate thorough consideration, given the sensitive nature of neural data. The establishment of regulatory frameworks designed to forestall any potential misuse will assume paramount importance as BCIs gain wider traction. In the forthcoming decades, the convergence of robotics and neurotechnology stands poised to revolutionize our interactions with machines and to amplify human capabilities in profound ways.

## Implantation technologies bring breakthroughs in medical rehabilitation

The implementation of BCI signal acquisition devices for therapeutic or rehabilitative purposes has gained acceptance among patients with amyotrophic lateral sclerosis (ALS) conditions [162,206]. These patients typically experience minimal changes in their living environment, thereby maximizing the potential benefits derived from the superior signal accuracy and low surgical risk offered by minimal-invasive implantation technology. This technology has demonstrated significant advancements in domains such as motor rehabilitation and disease intervention. While invasive implantation technology currently experiences limited employment within clinical practice, its capacity to procure high-precision, cell-level signals hold the promise of future medical breakthroughs.

To illustrate, implanted BCIs have the potential to empower ALS patients with complete paralysis to exert control over external devices such as computer cursors or robotic limbs solely through their cognitions of limb movement [207,208]. Although the technology remains imperfect, ongoing strides in neural decoding algorithms and the longevity and biocompatibility of implantable devices are propelling rapid advancements.

In the future, implanted BCIs could transcend their role in aiding communication and mobility among paralyzed patients. Furthermore, implanted BCIs could intervene in debilitating neurological disorders like epilepsy or Parkinson's disease [209,210]. In the context of epilepsy, real-time detection and mitigation of seizures through electrical stimulation could be achieved by recording from electrode arrays positioned in seizure-prone brain regions. In the case of Parkinson's disease, BCIs could capture signals from dysregulated areas and administer targeted stimulation to reinstate normal dynamics. The precise closed-loop modulation of pathological brain activity bears the capacity to substantially enhance outcomes for these conditions. In addition, implantable technology has potential applications in psychiatric disorders, such as depression.

Amidst these promising prospects, the adoption of invasive BCIs for medical applications necessitates rigorous validation of long-term safety and efficacy through comprehensive clinical trials. Additionally, the assurance of cybersecurity is imperative. Nevertheless, given the transformative restoration of function already demonstrated, implanted BCIs appear primed to extend medical capabilities and significantly enhance patient outcomes in the forthcoming years. With collaborative efforts across disciplines, these emerging neural technologies could potentially bestow millions grappling with neurological conditions with the restoration of health, autonomy, and quality of life.

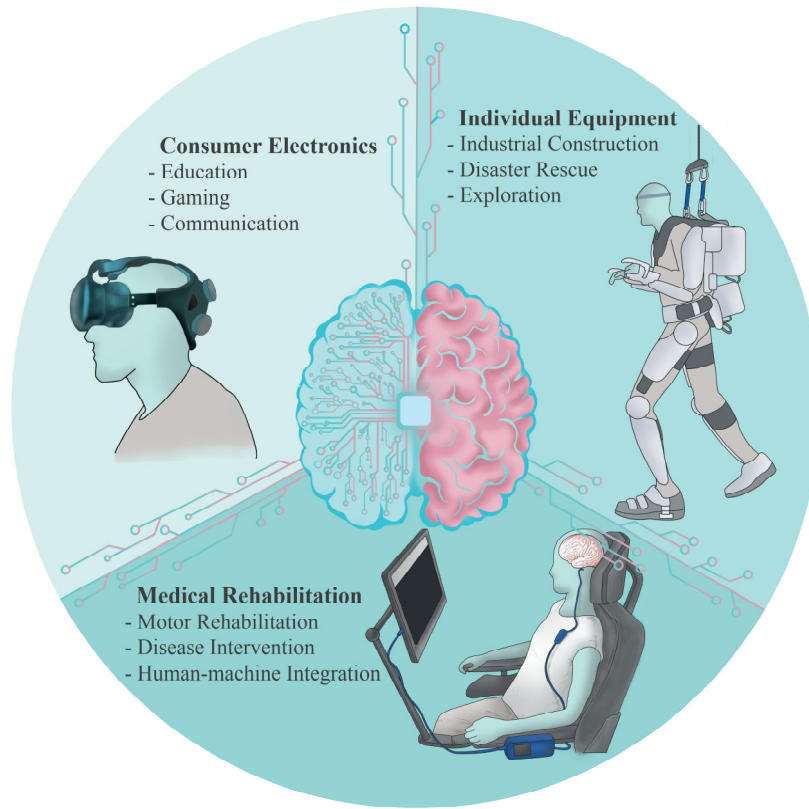

**Figure 7.** A prospective map of the future development of BCI technology. In the foreseeable future, non-invasive non-implantation technology has excellent room for development in consumer electronics, such as education, gaming, and communication. For individual equipment control in complex scenarios, there is more significant potential for signal acquisition technologies that provide stable signals with less trauma, including minimal-invasive non-implantation, intervention, and implantation technologies. For medical rehabilitation, minimal-invasive and invasive implantation technologies would bring more breakthrough possibilities.

## Acknowledgments

This work is supported by the National Natural Science Foundation of China (U2241208, 62171473, 61671424), the National Key Research and Development Program of China (2022YFC3602803). Authors would like to thank Ziyu Zhang from Xiamen University and Yuqing Zhao from the Central Academy of Fine Arts for their help in drawing the pictures in this article. The authors would also like to thank Prof. Guo Liang for his guidance in writing this paper.

## Competing interests

The authors declare no competing interests.